\documentclass[pdflatex,sn-mathphys-num]{sn-jnl}% Math and Physical Sciences Numbered Reference Style 
\usepackage{graphicx}%
\usepackage{multirow}%
\usepackage{amsmath,amssymb,amsfonts}%
\usepackage{amsthm}%
\usepackage{mathrsfs}%
\usepackage[title]{appendix}%
\usepackage{xcolor}%
\usepackage{textcomp}%
\usepackage{manyfoot}%
\usepackage{booktabs}%
\usepackage{algorithm}%
\usepackage{algorithmicx}%
\usepackage{algpseudocode}%
\usepackage{listings}%
\usepackage{cleveref}
\usepackage{hyperref}
\usepackage{multirow}
\usepackage{adjustbox}

\newcommand{\valstd}[2]{#1 {\footnotesize $\pm$ #2}}
\begin{document}

\title[Article Title]{Community Aware Temporal Network Generation}


\author*[1,2]{\fnm{Nicolò Alessandro} \sur{Girardini}}\email{ngirardini@fbk.eu; nicolo.girardini@unitn.it}

\author*[2]{\fnm{Antonio} \sur{Longa}}\email{antonio.longa@unitn.it}

\author[2]{\fnm{Gaia} \sur{Trebucchi}}

\author[3]{\fnm{Giulia} \sur{Cencetti}}

\author[2]{\fnm{Andrea} \sur{Passerini}}

\author[1]{\fnm{Bruno} \sur{Lepri}}

\affil[1]{\orgdiv{Mobile and Social Computing Lab (MobS)}, \orgname{Fondazione Bruno Kessler (FBK)}, \city{Trento}, \country{Italy}}

\affil[2]{\orgdiv{Dept. of Information Engineering and Computer Science (DISI)}, \orgname{University of Trento}, \city{Trento}, \country{Italy}}

\affil[3]{\orgname{Aix-Marseille Université, Université de Toulon, CNRS, CPT}, \city{Marseille}, \country{France}}


\abstract{
The advantages of temporal networks in capturing complex dynamics, such as diffusion and contagion, has led to breakthroughs in real world systems across numerous fields. In the case of human behavior, face-to-face interaction networks enable us to understand the dynamics of how communities emerge and evolve in time through the interactions, which is crucial in fields like epidemics, sociological studies and urban science. However, state-of-the-art datasets suffer from a number of drawbacks, such as short time-span for data collection and a small number of participants. Moreover, concerns arise for the participants' privacy and the data collection costs. Over the past years, many successful algorithms for static networks generation have been proposed, but they often do not tackle the social structure of interactions or their temporal aspect.

In this work, we extend a recent network generation approach to capture the evolution of interactions between different communities. Our method labels nodes based on their community affiliation and constructs surrogate networks that reflect the interactions of the original temporal networks between nodes with different labels. This enables the generation of synthetic networks that replicate realistic behaviors. We validate our approach by comparing structural measures between the original and generated networks across multiple face-to-face interaction datasets.
}


\maketitle

\section{Introduction}
\label{sec:intro}

The rise of network science in recent times fostered a great amount of innovation in multiple and diverse research fields~\cite{newman2010,barabasi2016}, from classical ones, such as biology~\cite{arrell2010network, kovacs2019network}, psychology~\cite{beard2016network, borsboom2021network}, economics~\cite{jackson2008}, to newer ones, such as urban science~\cite{batty2013,alessandretti2020scales,mauro2022generating, bontorin2024emergence} and neuroscience~\cite{de2014graph,bassett2017network}.
Moreover, network science naturally provides tools to understand how humans behave and interact among themselves on many levels, e.g., temporal level and group (or community) level~\cite{lazer2009computational,peixoto2017modelling,gelardi2021temporal}. In fact, it is easy to associate nodes with individuals and links between them with an interaction that occurs, or some other possible tie, such as being relatives.

Studying human interactions, specifically face-to-face or proximity interactions, became particularly relevant with the use of sensors that could easily be worn or with the use of smartphones. In this way, study participants could carry individual devices during their daily activities, allowing the devices to record their behavior in a natural setting. These studies are defined \textit{Living Labs}~\cite{eagle2006reality,cattuto2010, aharony2011social,isella2011pediatric,
stopczynski2014measuring,centellegher2016mobile,nepal2022covid} and proved to be very useful to study human interactions and their behavioral routines~\cite{eagle2009eigenbehaviors,starnini2017robust,ozella2019wearable,iacopini2022group, girardini2023adaptation,lotito2024multiplex,arregui2024patterns}. 
Nevertheless, the problems of recruitment and keeping the engagement high are significant for this kind of data collection, and resulting data are often limited in population size and temporal length.
These limitations affect the possibility of using the corresponding temporal networks for simulating realistic dynamics, such as epidemic spreading or opinion dynamics.

A possible solution to this problem is represented by the use of synthetic networks, that are generated to reproduce specific characteristics of real networks~\cite{pereira2020syntgen,zeno2021dymond,zeno2023dyane,cuppers2024flowchronicle}. Among the synthetic networks that can be built, surrogate networks, i.e., synthetic networks generated to imitate a specific input network, are particularly relevant in this task. In fact, they can be used as suitable replacements to the original real network and, if the generation model allows it, the surrogate networks are not bound to have the same temporal length and population size of the networks they are imitating~\cite{laurent2015calls,purohit2018temporal,cencetti2024generating}.
The model we propose possesses these characteristics, enabling the generation of realistic synthetic datasets without constraints on temporal length or network size.

We build our methodology upon a pre-existing method, namely \textit{ETN-gen}~\cite{longa2024generating}, which generates surrogates that reproduce the local behavior of nodes.
The idea of this generation process is to first decompose the original network into small temporal subnetworks, the Egocentric Temporal Neighborhoods (ETN), that encode the local neighborhood of each node on a short time-scale. The ETN can then be used as building blocks to build the surrogate networks.
A limitation of the \textit{ETN-gen} method is that it does not account for the heterogeneity of nodes. As a result, the behavior of all nodes in the generated network, while stochastically distinct, is derived from the same underlying patterns shared by all nodes. However, real-world networks often exhibit a modular structure. In social environments, the interplay between individuals with different roles or characteristics was shown to play a crucial role in shaping their distinct behaviors and interactions \cite{guimera2003self, hoffman2020model}.
This is true for the temporal component as well, as frequency of interactions and duration change according to the described interplay \cite{stopczynski2015temporal, sekara2016fundamental}.
An example is the Reality Mining behavioral study \cite{eagle2009eigenbehaviors}, which showed different interaction patterns for economics students, first year students and senior students.
Other environments also share these differences. For example, in a hospital setting, interactions among patients, nurses, and medical doctors can vary significantly due to the differing nature of their relationships (e.g., nurses tend to have only short exchanges with medical doctors while spending more time with patients~\cite{vanhems2013estimating}).

Therefore, we argue that limitations in characterizing the communities to which interacting individuals belong, represent a critical issue that must be addressed. To this end, we propose an enhancement of ETN-gen by introducing node labels that characterize nodes either based on their metadata, reflecting inherent characteristics when available, or by partitioning them into communities. As a result, a node’s ETN will encode not only the structure of its neighborhood, but also information about its community affiliation, as represented by these labels. This refinement is reflected in the generated networks, which exhibit a community organization similar to that of the original network.

Given the scope of this method, we evaluate it by comparing the structural metrics, meaningful for communities, of the starting network, with the generated ones. We do so both at an aggregated level and at the snapshot level, comparing how, in time, the metrics evolve in both networks. We choose several datasets from the \textit{SocioPatterns} study (see \Cref{subsec:data}) to evaluate our method. Our results show that our methodology is capable of reproducing networks similar to the original ones, significantly improving on the capabilities of the ETN-gen method~\cite{longa2024generating} when it comes to the interactions of individuals belonging to different communities and playing different roles. We believe this methodology has broad applicability across various research domains, including epidemiology, urban science, and the study of human interactions in confined settings, such as schools or workplaces.

\section{Materials and Methods}
\label{sec:methods}

Our methodology, \emph{Labeled ETN-gen} (LETN), builds upon the ETN-gen method, presented in \cite{longa2024generating}, to account for the interactions that happen among and across communities. We see this network of interactions as a temporal network, in which every snapshot or layer captures all the interactions (links) between study participants (nodes) in a specified interval. We call this interval \textit{gap} and it can be set depending on the type of interactions we have. For example, if the dataset contains face-to-face interactions we can identify meaningful layers grouping interactions with a 5 minutes interval, while if we take phone call interactions, a whole day would be a more appropriate interval.

As mentioned in the Introduction, we expand ETN-gen by adding a label attribute to the nodes present in the network. In this way, we are able to account for interactions within and between different types of nodes, e.g., students of different classes in a school. In the following subsections we describe the whole methodology and highlight where we applied specific changes to capture roles and communities. A visual illustration of the approach is shown in \Cref{fig:label_neigh} and \Cref{fig:generation}.

\subsection{Identifying Patterns}
\label{subsec:patterns}

Before the generation of the surrogate networks, we identify repeating patterns of the original network to be reproduced, so that the surrogate ones show the same properties.

\begin{figure}[t]
    \centering
    \includegraphics[width=\linewidth]{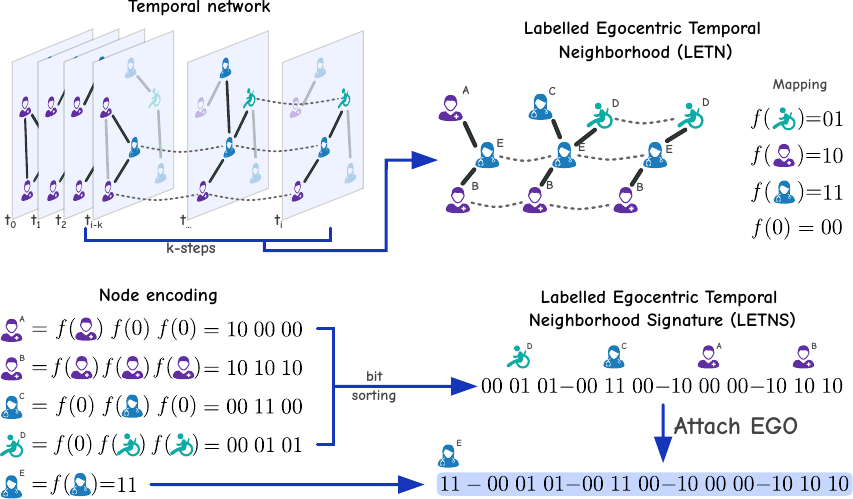}
    \caption{Labeled Egocentric Temporal Neighborhood Signature (LETNS) encoding. From a window of \textbf{k} steps of the original temporal network, we identify the temporal neighborhood of each node from an egocentric perspective. When encoding the signature, differently from the original approach (Egocentric Temporal Neighborhood), we use a binary mapping with $1$ when a link from the ego to another node is present in the current snapshot, and $0$ when it is not. Finally, we sort the temporal signatures of the neighbors and prepend the resulting string with the node encoding of the ego node to get the complete signature.}
    \label{fig:label_neigh}
\end{figure}

\paragraph{Labeled Egocentric Temporal Neighborhood (LETN)}
To encode meaningful patterns of interaction, we view networks from an egocentric perspective, that closely follows the evolution of interactions of a specific individual. We follow \cite{longa2022efficient,longa2024generating} by firstly identifying an \textit{Egocentric Temporal Neighborhood (ETN)} as all the links the ego node forms in a window of subsequent layers. The window size is identified with a parameter \textit{k}, for which an optimal value of $2$ is obtained in \cite{longa2024generating}. As in the previous work, we encode the neighborhood in a unique signature. However, here, nodes are represented using a mapping function encoding node labels into unique binary strings, the {\em node encoding}. For each node that is connected to the ego node at least once in the window, we represent its temporal signature by concatenating its node encoding in the snapshots where it is connected to the ego node, and using 0 for the snapshots where the link is absent. The following step involves sorting all the temporal signatures of the neighbors in a lexicographic order, so to eliminate dependency on the order in which neighbors are processed. Finally, we add the node encoding of the ego node at the beginning of the string, completing the \textit{Labeled Egocentric Temporal Neighborhood Signature (LETNS)}. \Cref{fig:label_neigh} illustrates this first step, by showing an example of interaction between a doctor (the ego E), two nurses (A and B), another doctor (C) and a patient (D).

\paragraph{Probability Dictionaries}
We then use the LETNS to build a dictionary of egocentric temporal neighborhood extensions. Following the original method, for each LETNS we mask the bits corresponding to last temporal layer, i.e., for each neighboring node we only consider its $k-1$ encodings. From the example in \Cref{fig:label_neigh}, the signature 11 000101 101010, becomes 11 0001x 1010y when masked. These masks are used as keys in a dictionary of temporal extension probability distributions, in which we will store each of the possible extensions, using their counts to compute their distribution. In performing this operation, the original method also uses a \textit{local split} and a \textit{global split}, which allow the generation to capture the periodicity in time. In fact, the method divides all the snapshots in smaller local splits, which repeat over global splits. In most use cases this means considering daily moments, e.g., hours (local split), that repeat across days (global split). Therefore, by grouping all the observations of the local splits across global splits to compute the local distribution, the method stores a dictionary for each local split. In this way we can capture how the individuals have different interactions during the day, and we can also reproduce the periodicity across the global split. Thus, this step is able to model the preferences of nodes with a certain label to interact with nodes with specific labels, also taking into account the time of the interactions.

\subsection{Surrogate Network Generation}
\label{subsec:generation}

Once the recurring patterns over local and global splits are identified, it is possible to generate the surrogate networks, following the original methodology.

\begin{figure}[ht]
    \centering
    \includegraphics[width=\linewidth]{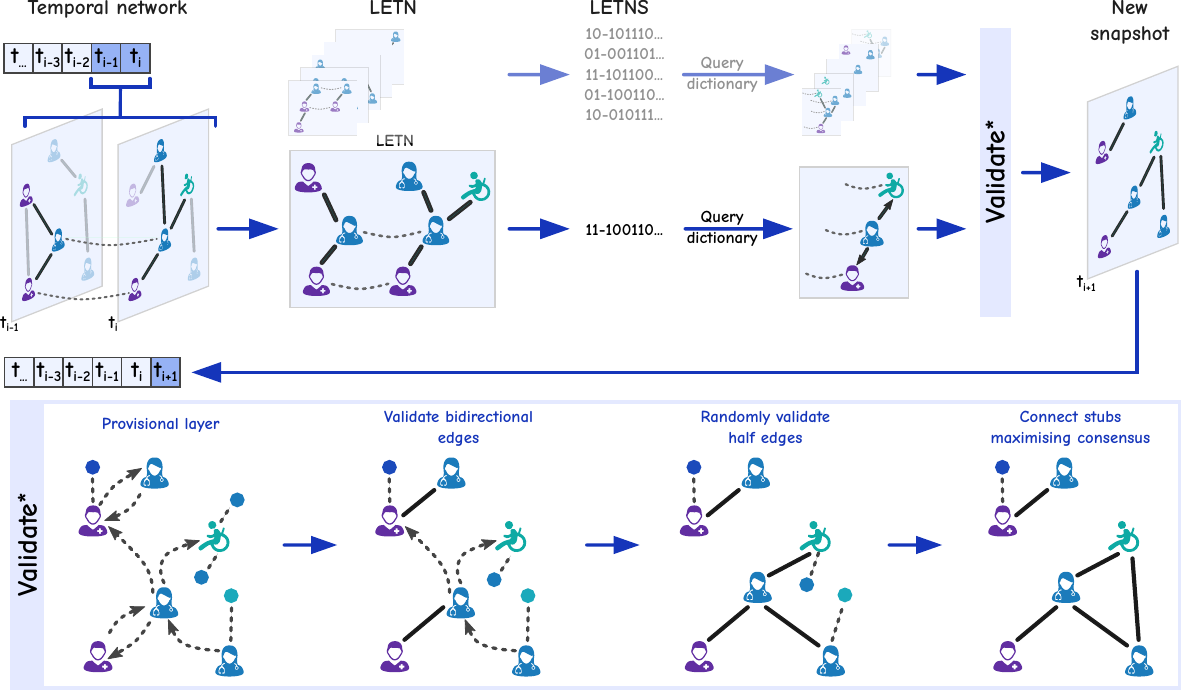}
    \caption{Generation step for the surrogate network. Starting from $k-1$ graphs as seeds, for each node we consider its LETN and compute the signature, which is used to sample from the probability dictionaries the desired connections in the new layer for that node. After sampling the dictionaries for each of the egos, we validate our \textit{provisional layer} with the displayed rules, which aim at replicating the interactions across communities present in the original network. The new layer can then be added to the temporal network and be used for the generation of the following step.}
    \label{fig:generation}
\end{figure}

\paragraph{Provisional Layer}
The generation steps follow again the original method, but by using the label-dependent dictionaries computed in the previous step. As displayed in \Cref{fig:generation}, the process starts by using $k-1$ layers as seeds, and extracting the LETNS for each node of the network. In this way, to generate the edges of the following snapshot, or layer, we can sample, with the current neighborhood signature, from the probability dictionaries defined above to obtain layer $k$ based on the possible extensions. This will determine a \textit{provisional layer}, in which each node has some desired connections to nodes with specific labels. Once the provisional layer has been validated (see next paragraph), the new snapshot is generated, added in the new network, and used for the generation of the following step. The steps can be repeated until the desired length of the network is reached.

\paragraph{Connection Validation}
The validation step applies a series of rules needed to reach a consensus among the links required by all the ego nodes when choosing their neighborhood. This step differs from \cite{longa2024generating} as the rules take into account the node labels. The validation, graphically shown in \Cref{fig:generation}, consists in:
\begin{enumerate}
    \item Validate the bidirectional edges, which represent the reciprocal requests (node $i$ requires a connections with node $j$ and node $j$ with node $i$). Since the two nodes agree to be connected the link is accepted. This maximizes the adherence of the final layer to the sampled LETN.
    \item Randomly validate half of the unidirectional edges, i.e., the unreciprocal requests (node $i$ requires a connection with node $j$ but node $j$ does not require the connection with node $i$).
    \item Connect the remaining stubs maximizing the consensus, i.e., maximizing the matching of the remaining reciprocal requests.
\end{enumerate}

\subsection{Label Absence}
\label{subsec:no_label}

If nodes labels are not provided as metadata with the temporal network dataset, we can assign labels based on the network community structure.

A first strategy consists in considering the aggregated network obtained by collapsing all the layers in one static network (where links are weighted according to the number of times that they appeared in the temporal layers), and to apply a community detection algorithm. The labels can then be assigned according to the obtained communities. We chose to use the Louvain algorithm \cite{blondel2008fast}, but other choices, like the Leiden algorithm \cite{traag2019louvain}, are possible. We call this strategy \textit{Community-LETN (CLETN)}.

A second strategy, which we call \textit{Dynamic-LETN (DLETN)}, is instead useful when the communities change significantly during the observation period. This happens, for instance, for the school datasets, since communities during classes are different than communities during breaks.
The strategy consists in aggregating at the local split level, thus obtaining multiple aggregated networks instead of only one. The community detection can then be applied to every aggregated network, obtaining different partitions for each local split. In this approach, the probability dictionaries will use these different partitions for each local split, resulting in different dictionaries for each local split.

\subsection{Datasets}
\label{subsec:data}

To ascertain the validity of the synthetic networks we generate, we select 7 datasets from the SocioPatterns repository (see Data Availability for details on how to access them). They contain data about face-to-face interactions over different period of times in different environments. Specifically, we select one dataset corresponding to a primary school \cite{stehle2011high, gemmetto2014mitigation}, three for a high school \cite{fournet2014contact, mastrandrea2015contact}, one for a hospital \cite{vanhems2013estimating} and two for an office workplace \cite{genois2015data, genois2018can}.
Each of these datasets contains metadata, setting the nodes label: there are classes and teachers for the schools, doctors and patients and nurses for the hospital, and office roles for the workplace datasets. This allowed us to test the base method and also the CLETN and DLETN extensions. The following \Cref{tab:datasets} shows descriptive statistics of the used datasets, in which we see how the chosen sample has different ratios of labels and participants and data collection duration and number of interactions.

\begin{table}[ht]
    \centering
    \caption{Statistical description of the datasets. The table lists number of participants, unique labels in the original metadata, length of the data collection (in days and hours) and total number of interactions.}
    \label{tab:datasets}
    \begin{tabular}{lcccc}
        \toprule
        \textbf{Dataset} & \textbf{Participants} & \textbf{Unique Labels} & \textbf{Length} & \textbf{Interactions}\\
        \midrule
        Primary School & 242 & 6 & 1d 9h & 125773 \\
        \midrule
        HighSchool 2011 & 126 & 4 & 3d 4h & 28561 \\
        \midrule
        HighSchool 2012 & 180 & 5 & 8d 11h & 45047 \\
        \midrule 
        HighSchool 2013 & 327 & 9 & 4d 5h & 188508 \\
        \midrule
        Hospital & 75 & 4 & 4d & 32424 \\
        \midrule
        Workplace 2013 & 92 & 5 & 11d 10h & 9827 \\
        \midrule
        Workplace 2015 & 217 & 12 & 11d 12h & 78249 \\
        \bottomrule
    \end{tabular}
\end{table}

In order to avoid possible biases in specific generations, for each of these datasets we generated 10 synthetic networks, starting from snapshots of the original graph as seeds, and then changing the seed for the dictionary sampling and validation step, and averaged the results. For these generations we used a window of $k=2$ and $gap=300$, i.e., snapshots of 5 minutes are used to group interactions.

\section{Results}
\label{sec:res}

\begin{figure}[t]
    \centering
    \includegraphics[width=1\linewidth]{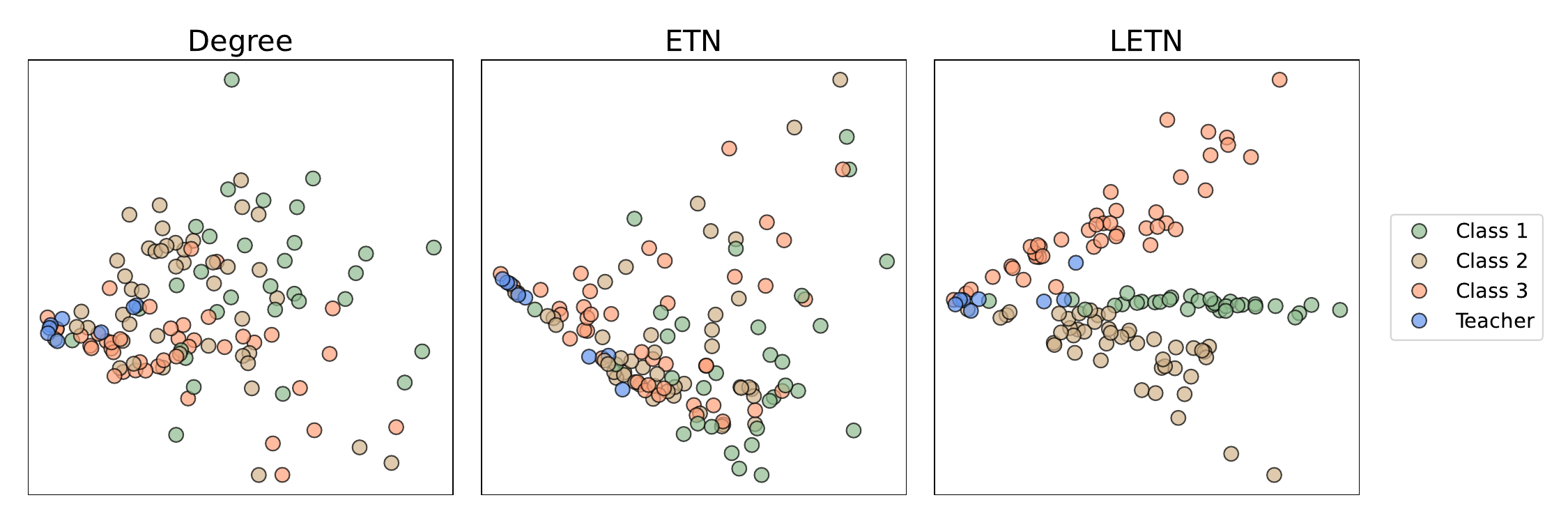}
    \caption{PCA visualization of the space of neighborhoods of the High School 2011 dataset. On the left we have the visualization of the degree counts via PCA, showing how differentiating nodes is not a trivial task. Comparing the visualizations for ETN and LETN generated networks, we see that the neighborhoods identified by ETN are not sufficient to distinguish clusters of nodes that have a similar behavior. With LETN, instead, we are able to construct clusters of the three classes.}
    \label{fig:pca_clusters}
\end{figure}

Our analyses aim to prove that the addition of labels in the ETN-gen methodology allows to capture the structure of interactions among and across communities, while the original ETN-gen approach misses these structures. Thus, the first step in this direction is visualizing how interaction neighborhoods encoded by LETN are more indicative of communities than the ones encoded by ETN. To do so, we identify unique neighborhood signatures, using both ETN signatures and LETN signatures, across all nodes and temporal layers for the High School 2011 original network. Then, for each node, we count the occurrence of each of the signatures found by the two methods, thus generating a vector per node. After this, we apply Principal Component Analysis (PCA) \cite{jolliffe2016principal} to visualize whether nodes of similar communities have similar structures of neighborhood. This is indeed the case, as shown in \Cref{fig:pca_clusters}. In the figure we firstly visualize, with the same technique, the degree of nodes, showing how the task is not trivial. Then, we show the outcome of PCA for the original network when the patterns are identified by either ETNS or LETNS. It is clear that, using LETNS, we can capture community patterns with the neighborhoods that it identifies, displayed by the fact that nodes belonging to the same class form clear clusters in the visualization. On the contrary, ETNS is not able to capture these dynamics. An interesting observation here, is that the behavior of teachers, who, differently from students, typically move across classes rather then interacting between themselves, is captured as well.

\begin{figure}[ht]
    \centering
    \includegraphics[width=0.99\linewidth]{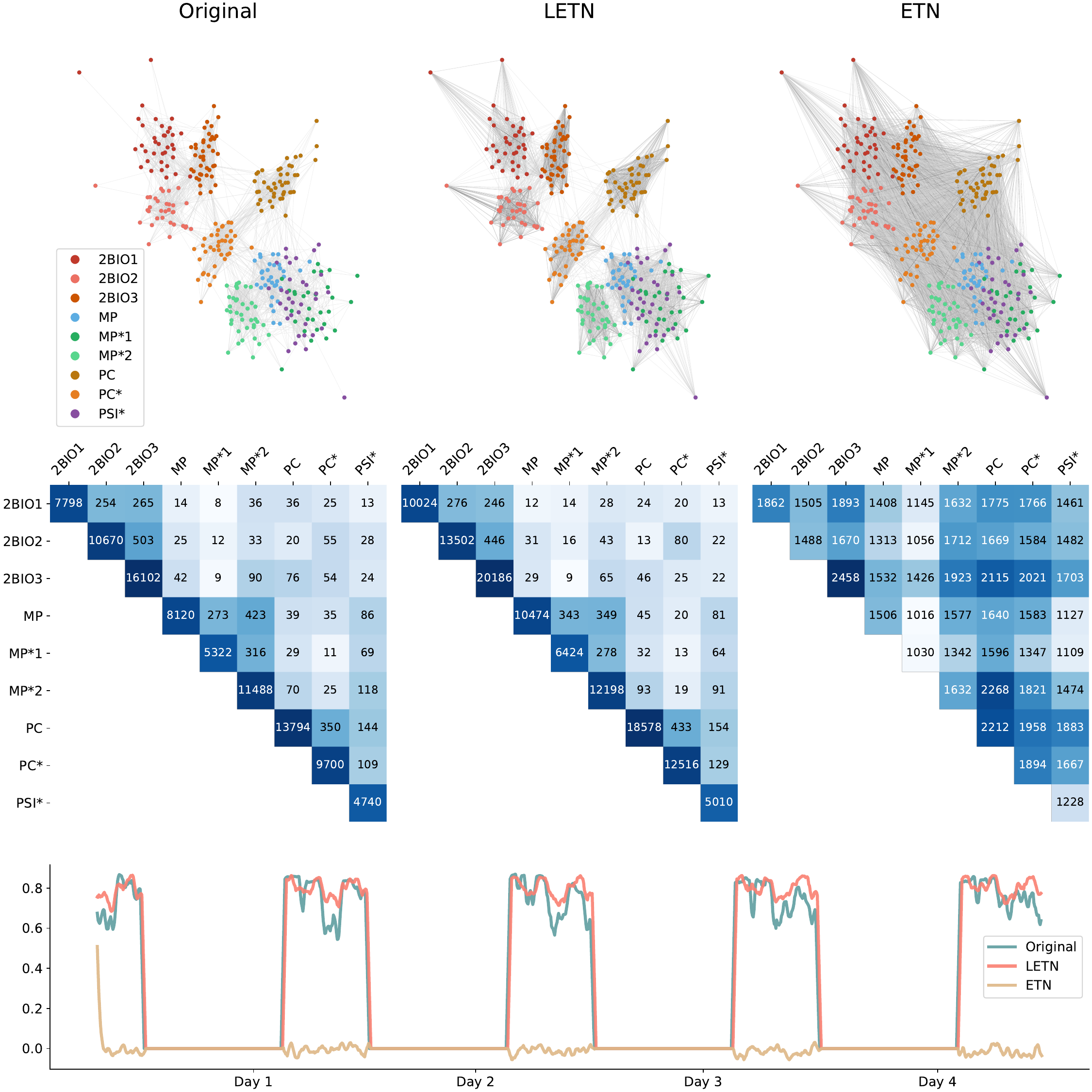}
    \caption{Comparison of the original network with networks generated by LETN and ETN respectively for the High School 2013 dataset. Panel \textbf{A} shows the networks aggregated over the whole measurement period. Our method generates a network with many interactions inside communities, and less outside communities, better resembling the original network. The ETN method, instead, generates random connections. These interactions are shown in the heatmaps of Panel \textbf{B}, where columns and rows represent different classes of the school. We see that the original patterns are reproduced by LETN, but not ETN. The final Panel \textbf{C} shows modularity through time (set to 0 when there were no interactions). Following the above observations, ETN has a poor modularity, while LETN closely follows the original network's trend.}
    \label{fig:hs13_res}
\end{figure}

Following this, we visualize the improvements on the base methodology on the High School 2013 dataset, as it presents more classes, and thus labels, to test the validity of the model. \Cref{fig:hs13_res} shows in three panels how the method we are proposing is able to capture communities better than the previous one. First of all, we show the original network, a LETN generated network and an ETN generated network, by aggregating all the temporal layers of the network, thus linking nodes that have at least one interaction during the measurement period. It is evident how networks obtained with the LETN method are characterized by a larger quantity of links inside the communities, resembling more the original network than ETN, where instead nodes are linked independently on the communities. The second row of \Cref{fig:hs13_res} shows more in detail how the interactions are distributed among and across communities, by using heatmaps (log-scaled colors). We also see, once again, how LETN is able to well reproduce how the interactions are distributed across communities. The ETN method, instead, is not able to capture any such pattern, as it is limited by not including labels in the neighborhood encoding. The final Panel \textbf{C}, shows the modularity (original labels are used), which is a proxy of the goodness of network partitions, over the measurement period. For visualization purposes, we set modularity to 0 when there were no interactions recorded, i.e., hours not at school. This step also shows the characteristic of the base method to capture interactions periodicity in the generated networks. We see that ETN is not able to generate a high modularity (the starting value is due to the seed graphs used in the generation), which always stays close to 0, meaning that links among and across communities are random and do not follow a fixed structure. Instead, LETN closely follows the trend of the original network's modularity.

\begin{figure}[t]
    \centering
    \includegraphics[width=0.99\linewidth]{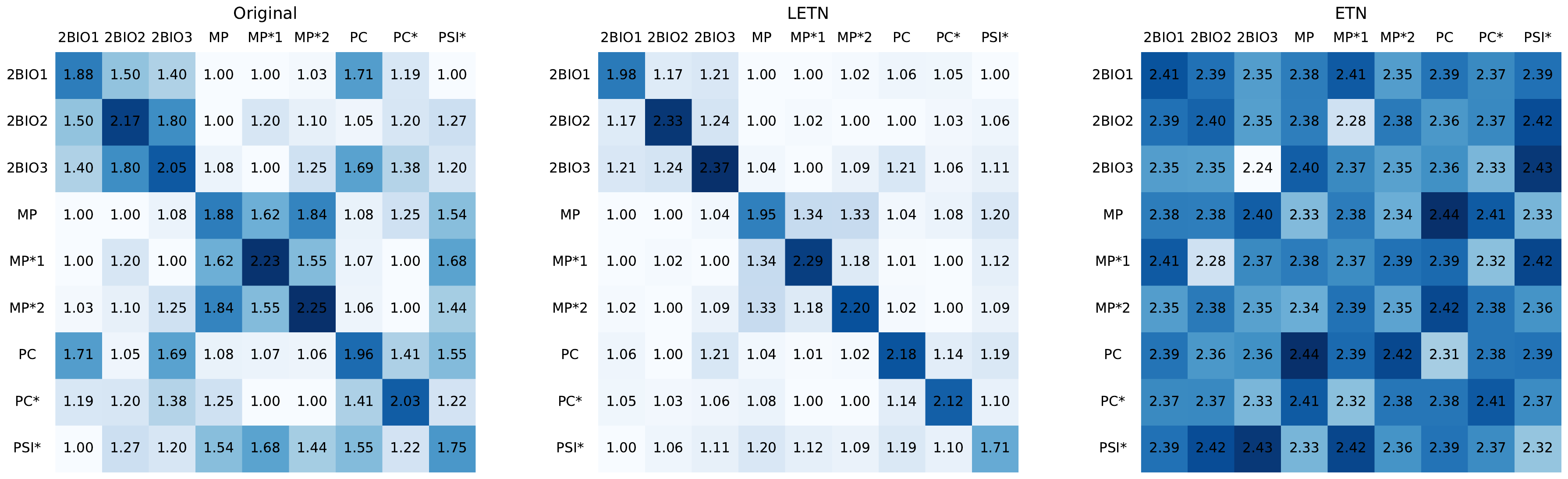}
    \caption{Mean number of consecutive interactions across communities, i.e., classes, of the High School 2013 dataset. The heatmaps show, on average, how many consecutive snapshots does an interaction between two communities last. We see how LETN, with the use of labels, can improve the ETN-based approach on reproducing the average duration of the face-to-face interactions.}
    \label{fig:mean_inter_time}
\end{figure}

Another improvement of the LETN method consists in the ability of our method to better reproduce the duration of the interactions among individuals of different communities. For this purpose, we have computed, for each interaction, the number of consecutive snapshots in which it is present. Then, we aggregate these computed lengths over the communities the nodes belong to, and average them, obtaining the mean duration of an interaction between nodes depending on their community. The result is shown in \Cref{fig:mean_inter_time}, which reports the average number of snapshots that an interaction between two specific communities lasts. We see an interesting behavior in the matrix representing the original dataset, namely that not only long interactions are present inside the same community (diagonal of the matrix), but that clusters are formed around classes that belong to the same subject. Therefore, it is meaningful that the LETN method is able to capture these patterns, while ETN, overestimating the number of interactions, also overestimates the average length of interactions and is not able to model any of these patterns.

\begin{figure}[t]
    \centering
    \includegraphics[width=0.99\linewidth]{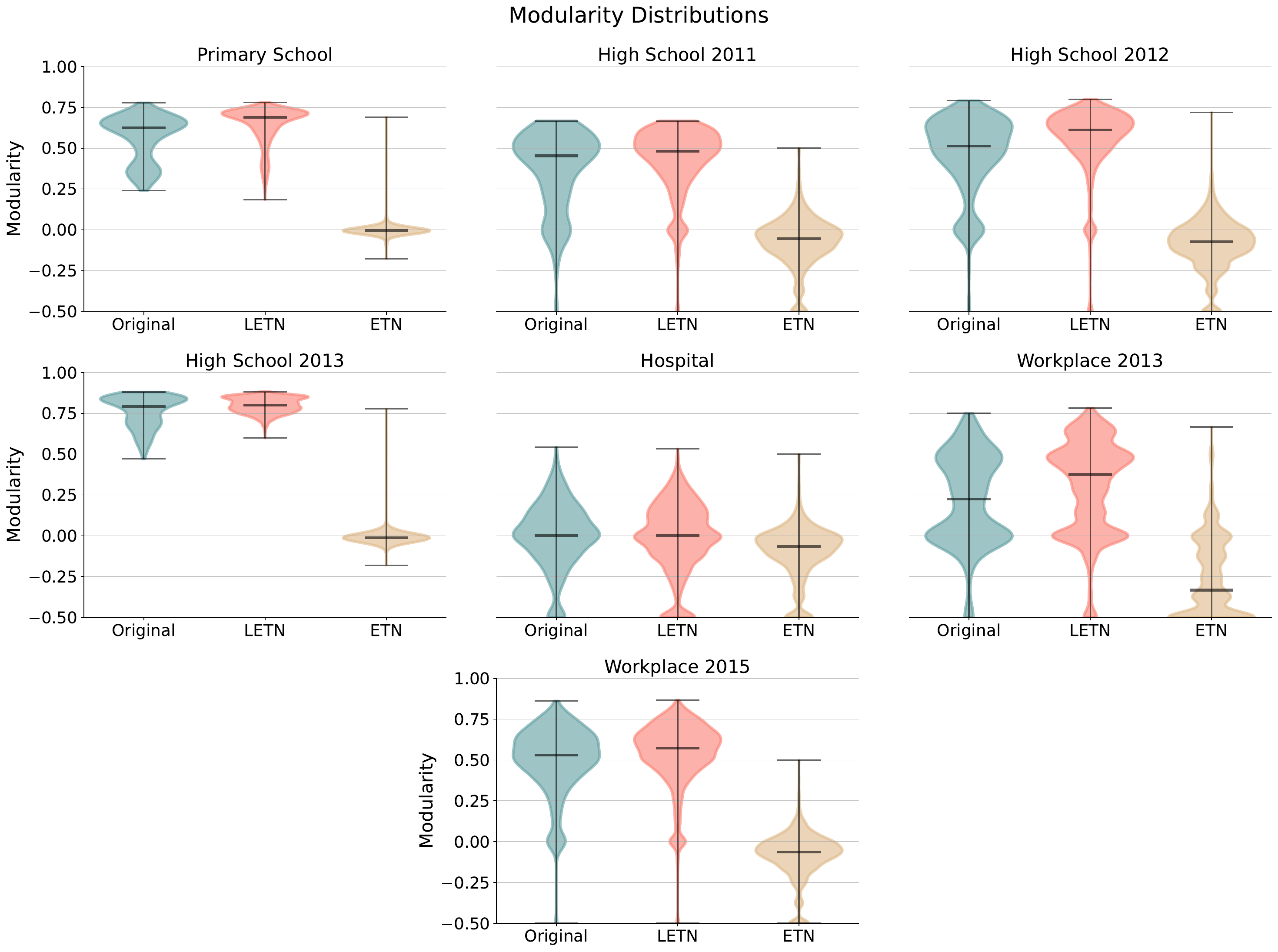}
    \caption{Distributions of modularity across all datasets, for the original network and the ones generated by LETN and ETN (the middle line is the median). We see how the generation of LETN is able to produce distributions very similar to the original ones, while ETN is not able to.}
    \label{fig:mod_distros}
\end{figure}

\begin{figure}[ht]
    \centering
    \includegraphics[width=0.99\linewidth]{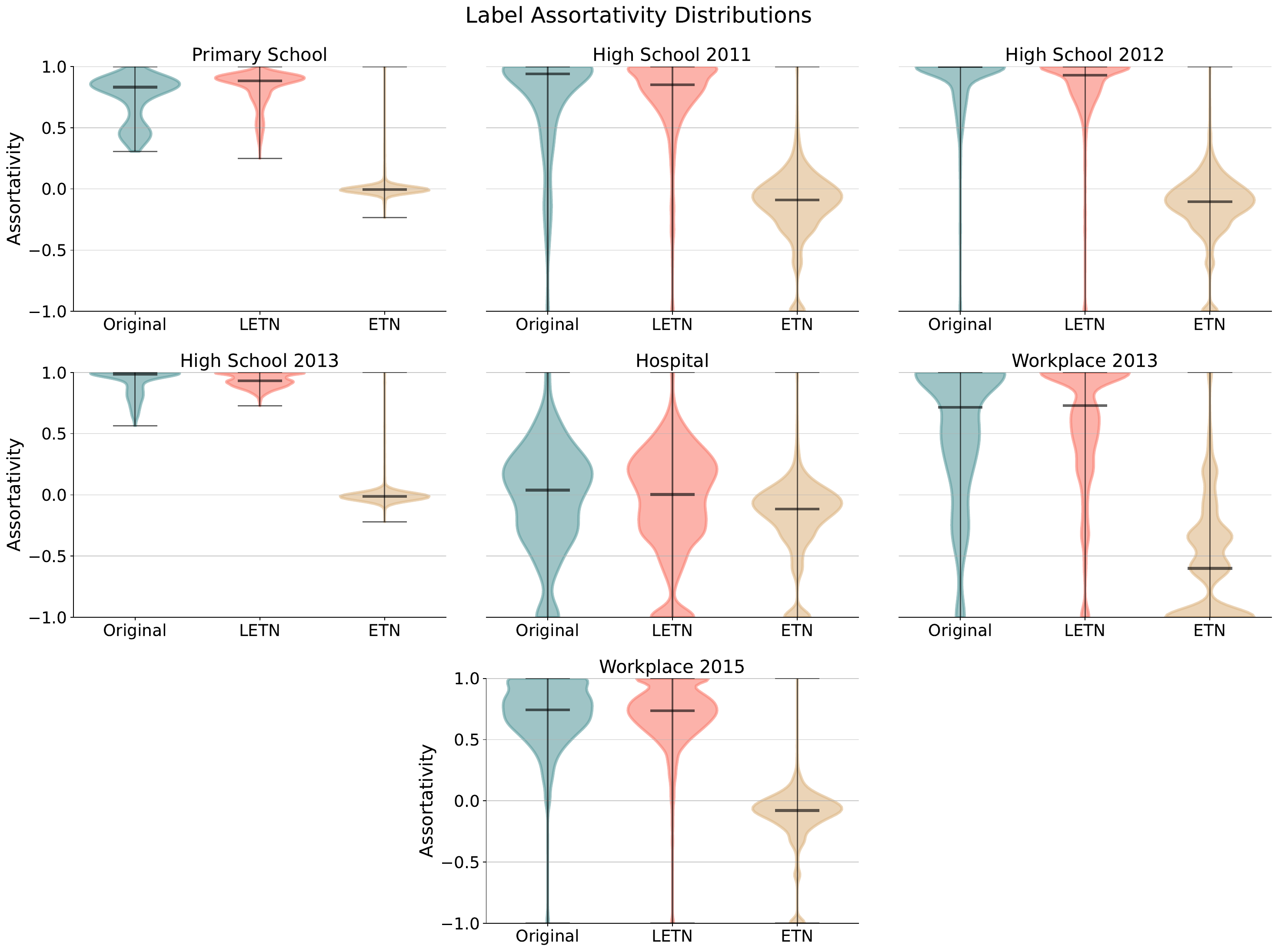}
    \caption{Distributions of label assortativity across all datasets, for the original network and the ones generated by LETN and ETN (the middle line is the median). We see how the generation of LETN is able to capture the interactions between nodes with different labels, while ETN cannot, as it does not encode labels}
    \label{fig:assort_distros}
\end{figure}

The outcomes presented for the High School 2013 dataset are similar to the ones that we obtain for all the other datasets we considered. \Cref{fig:mod_distros} and \Cref{fig:assort_distros} summarize this by representing the distribution of \textit{modularity} and \textit{label assortativity} for the original datasets, and for the synthetic networks generated by ETN and LETN. It is evident how the modularity distributions of the original datasets are well reproduced by the LETN method. Similar results are obtained for label assortativity. These results are a validation of how we are able to capture structures of interactions which depend on the type (e.g., role) of the nodes that participate in those interactions.

All the shown results correspond to $gap=5min$, which has resulted to be the most suited for the identification of interactions. However results do not change when we use  $gap=15min$, as reported in \Cref{tab:results_gap}, which shows the average modularity and label assortativity over the measurement period. In the same table we also report the results obtained for the aggregated networks (i.e., when all the temporal layers are collapsed to form one static network).

\begin{table}[ht]
    \centering
    \caption{\normalsize Average modularity and label assortativity of all the methods with gaps of 5 minutes, 15 minutes and the aggregated graph.}
    \label{tab:results_gap}
    \resizebox{\textwidth}{!}{\begin{minipage}{\textwidth}
        \begin{tabular}{llcccccc}
            \toprule
            \textbf{Dataset} & \textbf{Method} & \multicolumn{3}{c}{\textbf{Modularity}}& \multicolumn{3}{c}{\textbf{Label Assortativity}}\\
            & & 5min & 15min & Aggr & 5min & 15min & Aggr. \\
            \cmidrule(l{2pt}r{2pt}){1-1}
            \cmidrule(l{2pt}r{2pt}){2-2}
            \cmidrule(l{2pt}r{2pt}){3-5}
            \cmidrule(l{2pt}r{2pt}){6-8} 
            
            \multirow{3}{*}{\shortstack[l]{Primary\\School}} & Original & \valstd{0.56}{0.14} & \valstd{0.57}{0.14} & 0.59 & \valstd{0.74}{0.19} & \valstd{0.71}{0.20} & 0.31 \\
            & LETN & \valstd{0.64}{0.11} & \valstd{0.65}{0.10} & \valstd{0.69}{0.01} & \valstd{0.83}{0.13} & \valstd{0.65}{0.10} & \valstd{0.52}{0.01} \\
            & ETN & \valstd{0.00}{0.06} & \valstd{0.00}{0.04} & \valstd{0.66}{0.01} & \valstd{0.00}{0.08} & \valstd{0.00}{0.04} & \valstd{0.01}{0.00} \\
            \midrule
            \multirow{3}{*}{\shortstack[l]{HighSchool\\2011}} & Original & \valstd{0.37}{0.24} & \valstd{0.37}{0.25} & 0.56 & \valstd{0.75}{0.40} & \valstd{0.68}{0.43} & 0.65 \\
            & LETN & \valstd{0.42}{0.15} & \valstd{0.44}{0.14} & \valstd{0.56}{0.01} & \valstd{0.77}{0.19} & \valstd{0.44}{0.14} & \valstd{0.58}{0.01} \\
            & ETN & \valstd{-0.08}{0.08} & \valstd{-0.06}{0.06} & \valstd{0.48}{0.01} & \valstd{-0.13}{0.16} & \valstd{-0.06}{0.06} & \valstd{0.00}{0.01} \\
            \midrule
            \multirow{3}{*}{\shortstack[l]{HighSchool\\2012}} & Original & \valstd{0.46}{0.24} & \valstd{0.50}{0.20} & 0.66 & \valstd{0.89}{0.26} & \valstd{0.89}{0.21} & 0.63 \\
            & LETN & \valstd{0.55}{0.15} & \valstd{0.56}{0.14} & \valstd{0.67}{0.01} & \valstd{0.84}{0.16} & \valstd{0.56}{0.14} & \valstd{0.60}{0.01} \\
            & ETN & \valstd{-0.09}{0.07} & \valstd{-0.07}{0.06} & \valstd{0.55}{0.01} & \valstd{-0.14}{0.14} & \valstd{-0.07}{0.06} & \valstd{0.01}{0.00} \\
            \midrule 
            \multirow{3}{*}{\shortstack[l]{HighSchool\\2013}} & Original & \valstd{0.76}{0.10} & \valstd{0.77}{0.08} & 0.80 & \valstd{0.92}{0.11} & \valstd{0.91}{0.10} & 0.65 \\
            & LETN & \valstd{0.80}{0.04} & \valstd{0.80}{0.04} & \valstd{0.81}{0.01} & \valstd{0.93}{0.05} & \valstd{0.80}{0.04} & \valstd{0.62}{0.01} \\
            & ETN & \valstd{-0.01}{0.06} & \valstd{0.00}{0.06} & \valstd{0.77}{0.01} & \valstd{-0.01}{0.07} & \valstd{0.00}{0.06} & \valstd{0.01}{0.01} \\
            \midrule
            \multirow{3}{*}{Hospital} & Original & \valstd{0.00}{0.20} & \valstd{-0.01}{0.20} & 0.14 & \valstd{-0.03}{0.43} & \valstd{-0.08}{0.40} & -0.12 \\
            & LETN & \valstd{-0.02}{0.16} & \valstd{-0.02}{0.17} & \valstd{0.13}{0.01} & \valstd{-0.07}{0.35} & \valstd{-0.02}{0.17} & \valstd{-0.13}{0.01} \\
            & ETN & \valstd{-0.11}{0.09} & \valstd{-0.10}{0.09} & \valstd{0.10}{0.01} & \valstd{-0.20}{0.21} & \valstd{-0.10}{0.09} & \valstd{-0.01}{0.01} \\
            \midrule
            \multirow{3}{*}{\shortstack[l]{Workplace\\2013}} & Original & \valstd{0.22}{0.28} & \valstd{0.33}{0.25} & 0.55 & \valstd{0.55}{0.56} & \valstd{0.65}{0.40} & 0.52 \\
            & LETN & \valstd{0.29}{0.12} & \valstd{0.40}{0.11} & \valstd{0.55}{0.01} & \valstd{0.60}{0.18} & \valstd{0.40}{0.11} & \valstd{0.41}{0.01} \\
            & ETN & \valstd{-0.25}{0.08} & \valstd{-0.16}{0.07} & \valstd{0.42}{0.01} & \valstd{-0.54}{0.14} & \valstd{-0.16}{0.07} & \valstd{-0.01}{0.01} \\
            \midrule
            \multirow{3}{*}{\shortstack[l]{Workplace\\2015}} & Original & \valstd{0.50}{0.21} & \valstd{0.52}{0.20} & 0.57 & \valstd{0.71}{0.26} & \valstd{0.71}{0.26} & 0.32 \\
            & LETN & \valstd{0.54}{0.12} & \valstd{0.55}{0.13} & \valstd{0.62}{0.01} & \valstd{0.70}{0.10} & \valstd{0.55}{0.13} & \valstd{0.30}{0.01} \\
            & ETN & \valstd{-0.08}{0.08} & \valstd{-0.07}{0.07} & \valstd{0.40}{0.01} & \valstd{-0.13}{0.14} & \valstd{-0.07}{0.07} & \valstd{0.00}{0.01} \\
            \bottomrule
        \end{tabular}
    \end{minipage}}
\end{table}

Finally, we validate the two strategies proposed to deal with the absence of labeled data, namely CLETN, which assigns labels based on the partitions found on the aggregated graph, and DLETN, which assigns labels based on the partitions computed at the local split level. To do so, we compare the trend of modularity and label assortativity over all the snapshots between their generated networks and the original one. In particular, we compute \textit{modularity}, \textit{label assortativity} and \textit{clustering coefficient} for each snapshot of the original network and the generated ones, then, we compute a euclidean distance on the resulting vectors. In this way we compare how much the generation of both LETN, CLETN, and DLETN follow the trend of the original network when it comes to these structural measures. From the results in \Cref{tab:results_estensions}, it is clear that both the two proposed extensions do not differ significantly from the results of the main LETN methodology, thus proving to be valid alternatives when metadata with node characteristics are not available.

\begin{table}[ht]
    \centering
    \caption{Euclidean distance for the selected metrics for LETN and the CLETN and DLETN extensions. We compute the metrics at each snapshot for the original data and 10 generated networks per method, then we compute the euclidean distance between the resulting vectors, for which we report average and error in the table. The very close results confirm the validity of the extensions we propose for LETN.}
    \label{tab:results_estensions}
    \begin{tabular*}{\textwidth}{@{\extracolsep{\fill}} llccc}
        \toprule
        \textbf{Dataset} & \textbf{Method} & \textbf{Modularity} & \textbf{Label Assortativity} & \textbf{Clustering Coefficient}\\
        \midrule
        \multirow{3}{*}{\shortstack[l]{Primary\\School}} & LETN & \valstd{3.21}{0.05} & \valstd{4.07}{0.06} & \valstd{1.72}{0.02} \\
        & CLETN & \valstd{3.07}{0.05} & \valstd{3.93}{0.06} & \valstd{1.55}{0.01} \\
        & DLETN & \valstd{3.21}{0.06} & \valstd{4.06}{0.08} & \valstd{1.43}{0.02} \\
        \midrule
        \multirow{3}{*}{\shortstack[l]{HighSchool\\2011}} & LETN & \valstd{6.45}{0.17} & \valstd{11.78}{0.21} & \valstd{0.79}{0.03} \\
        & CLETN & \valstd{6.07}{0.17} & \valstd{11.14}{0.40} & \valstd{0.80}{0.02} \\
        & DLETN & \valstd{5.69}{0.20} & \valstd{10.69}{0.40} & \valstd{0.80}{0.02} \\
        \midrule
        \multirow{3}{*}{\shortstack[l]{HighSchool\\2012}} & LETN & \valstd{13.63}{0.10} & \valstd{20.17}{0.15} & \valstd{0.76}{0.01} \\
        & CLETN & \valstd{11.40}{0.14} & \valstd{18.28}{0.19} & \valstd{0.76}{0.01} \\
        & DLETN & \valstd{12.63}{0.23} & \valstd{19.61}{0.33} & \valstd{0.76}{0.01} \\
        \midrule 
        \multirow{3}{*}{\shortstack[l]{HighSchool\\2013}} & LETN & \valstd{3.81}{0.04} & \valstd{4.35}{0.05} & \valstd{1.11}{0.01} \\
        & CLETN & \valstd{3.87}{0.04} & \valstd{4.87}{0.05} & \valstd{1.11}{0.01} \\
        & DLETN & \valstd{3.64}{0.03} & \valstd{4.39}{0.04} & \valstd{1.11}{0.01} \\
        \midrule
        \multirow{3}{*}{Hospital} & LETN & \valstd{7.09}{0.14} & \valstd{13.79}{0.29} & \valstd{1.48}{0.01} \\
        & CLETN & \valstd{7.86}{0.18} & \valstd{15.09}{0.39} & \valstd{1.56}{0.01} \\
        & DLETN & \valstd{7.47}{0.18} & \valstd{14.41}{0.35} & \valstd{1.56}{0.01} \\
        \midrule
        \multirow{3}{*}{\shortstack[l]{Workplace\\2013}} & LETN & \valstd{15.01}{0.28} & \valstd{27.06}{0.41} & \valstd{0.36}{0.01} \\
        & CLETN & \valstd{14.26}{0.16} & \valstd{26.19}{0.25} & \valstd{0.36}{0.01} \\
        & DLETN & \valstd{14.25}{0.19} & \valstd{26.17}{0.38} & \valstd{0.36}{0.01} \\
        \midrule
        \multirow{3}{*}{\shortstack[l]{Workplace\\2015}} & LETN & \valstd{13.98}{0.27} & \valstd{19.34}{0.29} & \valstd{0.82}{0.01} \\
        & CLETN & \valstd{12.83}{0.32} & \valstd{18.74}{0.45} & \valstd{0.82}{0.01} \\
        & DLETN & \valstd{13.14}{0.28} & \valstd{19.43}{0.48} & \valstd{0.82}{0.01} \\
        \bottomrule
    \end{tabular*}
\end{table}

\section{Discussion}
\label{sec:discussion}

The method we propose allows us to generate temporal networks that serve as surrogates to observed real-world datasets.
Our approach builds on ETN-gen~\cite{longa2024generating}, by adding a mechanism of node labeling, that has as a main advantage to reproduce the organization of nodes in communities, i.e., groups of nodes highly connected among them and less connected to the other groups of nodes.
Communities can be observed in any kind of datasets~\cite{leskovec2008statistical} but they are particularly relevant for social networks, in some cases spontaneously emerging as in groups of friends or colleagues~\cite{isella2011s, genois2018can}, and in some cases imposed by the environment setting, as in schools where students are usually separated into classes~\cite{stehle2011high}.
The structure created by the communities is not poorly relevant since it has a major effect on the global phenomena that can be observed on a network.
Examples are represented by spreading phenomena, where the effect of communities is to enhance the local spreading while slowing down the global one~\cite{salathe2010dynamics,cencetti2024temporal}, thus resulting in a completely different diffusion pattern with respect to networks without a community structure. 
Similar effects can be observed in opinion dynamics~\cite{peng2023role}, information transfer~\cite{danon2008impact}, and synchronization~\cite{arenas2008synchronization}.

In general, the existence of structure in a network is what distinguishes it from a random network, hence adding communities to surrogate networks makes them more similar to the original ones and, thus, more realistic.

In our method, the community structure is not imposed as a global or meso-scale feature, but it emerges from the microscopic dynamics, being incorporated in the local development rules and in how each node chooses its neighbors.
The ego-centric perspective also allows us to distinguish among nodes belonging to different categories. If categories are characterized by different individual behaviors in the original network, these differences are naturally reproduced in the surrogate networks.
In the hospital dataset, for instance, we can reproduce the fact that nurses are characterized by a higher number of interactions, especially with patients and among themselves, while medical doctors have less interactions but with a longer duration.
This mechanism increases the nodes heterogeneity, a peculiarity of social networks that the previous method, ETN-gen, could not reproduce.

With our method we showed how an already efficient method like ETN-gen can be further improved by adding node labels.
This represents only a little increase in the algorithm complexity with significant positive implications.
By using it, we are in fact able to reproduce interactions with a higher fidelity to the original network, both considering the distribution of links among nodes and their temporal duration.

This method can be applied in the cases where communities are provided by metadata or not, as in the second case they can be obtained by a simple community detection algorithm~\cite{girvan2002community}, which can be applied to the aggregated network or even to each singular snapshot of the temporal network.

In conclusion, we believe that this lightweight method can be a useful tool for the study of human interactions, particularly in the case of face-to-face interactions.

\section*{Data Availability}

All the dataset used in the analysis are publicly available on the \textit{SocioPatterns} project website: \url{http://www.sociopatterns.org}

\section*{Acknowledgments}

G.C. acknowledges the support of the European Union’s Horizon research and innovation program under the Marie Skłodowska-Curie grant agreement No 101103026. A.L., A.P., and B.L. acknowledge the support of the MUR PNRR project FAIR - Future AI Research (PE00000013) funded by the NextGenerationEU.
B.L. acknowledges also the support of the PNRR ICSC National Research Centre for High Performance Computing, Big Data and Quantum Computing (CN00000013), under the NRRP MUR program funded by the NextGenerationEU.


\bibliography{bibliography}% common bib file

\end{document}